\documentclass[12pt]{article}
\usepackage{epsfig}
\usepackage{amsfonts,amssymb,amsmath,graphicx}
\def\beq{\begin{equation}}
\def\eeq#1{\label{#1}\end{equation}}
\def\eeqn{\end{equation}}

\def\beqa{\begin{eqnarray}}
\def\eeqa#1{\label{#1}\end{eqnarray}}
\def\eeqan{\end{eqnarray}}

\let\bar=\overbar

\def\Dslash{\not{\hbox{\kern-4pt $D$}}}
\def\dslash{\not{\hbox{\kern-2pt $\del$}}}

\def\msb{{\bar{\ssstyle M \kern -1pt S}}}

\def\Title#1{\begin{center} {\Large {\bf #1} } \end{center}}

\begin{document}

\Title{AdS String: Classical Solutions \\ and Moduli Dynamics}

\bigskip\bigskip


\begin{raggedright}  

{\it Antal Jevicki \footnote{Based on talks given at ``Tenth workshop on non-perturbative Quantum Chromodynamics'', l'Institut Astrophysique de Paris, France, June 8-12, 2009; ``Summer school on AdS/CFT correspondence and its applications'', Tihany, Hungary, August 24-28, 2009.}, Kewang Jin\\
Department of Physics\\
Brown University, Box 1843\\
Providence, RI 02912, USA}
\bigskip\bigskip
\end{raggedright}

\section{Introduction}

In what follows we present a review of recently developed techniques for generating classical solutions of strings in Anti-de-Sitter space and for studying their classical dynamics. The initial arguments for the relevance of semiclassical calculations in AdS spacetime were given by Gubser, Klebanov and Polyakov \cite{Gubser:2002tv} who described the rotating, folded strings and the associated Yang-Mills operators. A generalization to the N-spike configuration was given by Kruczenski \cite{Kruczenski:2004wg}. In a rotating frame, the original configuration of \cite{Kruczenski:2004wg} was static, as such it corresponds to a ground state of a more general dynamical solution. Its study and full specification is of definite interest.

The spike configurations are associated with Minkowski worldsheets, they provide information about the semiclassical spectrum of the AdS string. The recently developed Wilson loop/Scattering amplitude duality requires classical solutions with Euclidean worldsheets which enter the construction of minimal area surfaces with polygonal boundary conditions \cite{Alday:2007hr}. The reduction technique that we have developed for the AdS problem is seen to apply both to Minkowski and Euclidean cases \cite{Jevicki:2007aa, Jevicki:2008mm, Jevicki:2009uz, Jevicki:2009bv}. It is based on a reduction originally introduced by Pohlmeyer \cite{Pohlmeyer} for nonlinear sigma models and applied by de Vega and Sanchez \cite{de Vega} to strings in de-Sitter spacetime. The reduction to the invariant fields provides a bridge to integrable equations of sinh-Gordon and more generally Toda type. These equations have soliton type solutions and can be studied through the inverse scattering method \cite{AKNS}.

One can see in \cite{Jevicki:2007aa, Dorey:2008vp} that a soliton configuration with a singularity at its location translates into a spike at the string level. This connection turns out to be most fruitful for the construction of dynamical multi-spike solutions. We summarize the construction given in \cite{Jevicki:2009uz} where a string solution associated with the most general N-soliton solution was given. We describe here in some detail the particular cases corresponding to one and two solitons and evaluate the energies. These provide information on the fluctuation spectrum of the AdS string and more generally its moduli dynamics.

It is useful to draw a parallel with an analogous construction found for the case of strings moving on $R \times S^n$. In the $n=2$ case one had the association (due to Hofman and Maldacena \cite{Hofman:2006xt}) between solitons of sine-Gordon and magnons of the string. In \cite{Aniceto:2008pc} the moduli space description was presented, given by the N-body problem of Calogero (or Ruijsenaars-Schneider type \cite{Ruijsenaars}). Our present work implies an analogous description in the case of $AdS_3$.

Regarding Euclidean worldsheets, the Pohlmeyer reduction turned out to be very useful for construction of minimal surfaces relevant to the Alday-Maldacena construction \cite{Alday:2009ga, Alday:2009yn}. We review here a simple configuration based on the sinh-Gordon soliton which translates into a solution originally constructed by Berkovits and Maldacena \cite{Berkovits:2008ic} representing a cusp near the boundary of AdS space, which can be simulated by a Higgs field cutoff in Yang-Mills theory \cite{Alday:2009zm}. For the problem of constructing minimal surfaces in AdS it turns out that one needs not soliton but instanton type solutions of sinh-Gordon and more generally Toda equations. Contrary to solitons no exact techniques exist for the instanton case. We describe an approximate series method for construction of such solutions \cite{Jevicki:2009bv}.

The content of this review is as follows. In section two, we review the Pohlmeyer reduction of classical strings in $AdS$. In section three, we present the most general N-spike solution in $AdS_3$ associated with N solitons of sinh-Gordon. In the concrete examples of one and two solitons, we exhibit explicitly the corresponding one and two spike dynamical solutions. For these we show the relationship between soliton energies and the corresponding string energies and momenta. In section four, we discuss an interesting correspondence between solitons and cusps with Euclidean worldsheets. Finally, we give a series solution to the $B_2$ Toda equations which are relevant to the Pohlmeyer reduction in $AdS_4$.

\section{Pohlmeyer reduction}

In general, sting equations in $AdS_d$ spacetime (in conformal gauge) are described by the non-compact nonlinear sigma model on $SO(d-1,2)$. Defining the $AdS_d$ space as $Y^2=-Y_{-1}^2-Y_0^2+Y_1^2+\cdots+Y_{d-1}^2=-1$, the action reads
\begin{equation}
S={\sqrt{\lambda} \over 4\pi}\int d\tau d\sigma \Bigl(\partial Y \cdot \partial Y + \Lambda (\sigma ,\tau)(Y \cdot Y+1)\Bigr),
\end{equation}
where $\tau,\sigma$ are the Minkowski worldsheet coordinates, the equations of motion are
\begin{equation}
\partial \bar{\partial} Y-(\partial Y \cdot \bar{\partial} Y) Y=0,
\label{stringsigma}
\end{equation}
with $z=(\sigma-\tau)/2$, $\bar{z}=(\sigma+\tau)/2$ and $\partial=\partial_\sigma-\partial_\tau$, $\bar{\partial}=\partial_\sigma+\partial_\tau$. In addition to guarantee the conformal gauge we have to impose the Virasoro conditions
\begin{equation}
\partial Y \cdot \partial Y = \bar{\partial} Y \cdot \bar{\partial} Y=0.
\label{Virasoro}
\end{equation}

It was demonstrated a number of years ago (by Pohlmeyer \cite{Pohlmeyer}) that nonlinear sigma models subject to Virasoro type constraints can be reduced to integrable field equations of sine-Gordon (or Toda) type. This reduction is accomplished by concentrating on $SO(d-1,2)$ invariant sub-dynamics of the sigma model. The steps of the reduction were well described in \cite{de Vega, Jevicki:2007aa} and consist in the following. One starts by identifying first an appropriate set of basis vectors for the string coordinates
\begin{equation}
e_i=(Y,\partial Y,\bar{\partial} Y,B_4,\cdots,B_{d+1}), \qquad i=1,2,\cdots,d+1,
\end{equation}
where $B_i$ form an orthonormal set $B_i \cdot B_j=\delta_{ij},B_i \cdot Y=B_i \cdot \partial Y=B_i \cdot \bar{\partial} Y=0$. Defining the scalar field $\alpha$ and two sets of auxiliary fields
\begin{eqnarray}
\alpha(z,\bar{z}) &\equiv& \ln[\partial Y \cdot \bar{\partial} Y], \\
u_i &\equiv& B_i \cdot \bar{\partial}^2 Y, \\
v_i &\equiv& B_i \cdot \partial^2 Y,
\end{eqnarray}
one can derive the equations of motion
\begin{eqnarray}
&&\partial \bar{\partial} \alpha-e^{\alpha}-e^{-\alpha}\sum_{i=4}^{d+1} u_i v_i=0, \\
&& \quad \partial u_i=\sum_{j \neq i}(B_j \cdot \partial B_i) u_j, \\
&& \quad \bar{\partial} v_i=\sum_{j \neq i}(B_j \cdot \bar{\partial}B_i) v_j.
\end{eqnarray}

In the case of $AdS_3$, there is only one vector $B_4$ so that $\partial u=0$, $\bar{\partial} v=0$. Therefore, the equation of motion for the scalar field $\alpha$ is simplified to be
\begin{equation}
\partial \bar{\partial} \alpha-e^{\alpha}-e^{-\alpha}u(\bar{z})v(z)=0.
\end{equation}
This is the generalized sinh-Gordon equation with two (anti)-holomorphic functions $u(\bar{z})$ and $v(z)$. In order to get the standard sinh-Gordon equation, we first shift the field
\begin{equation}
\alpha(z,\bar{z})=\hat{\alpha}(z,\bar{z})+\ln \sqrt{-u(\bar{z})v(z)},
\end{equation}
and then do a (conformal) change of variables
\begin{equation}
d\bar{z}'=\sqrt{2u(\bar{z})}d \bar{z}, \qquad d z'=\sqrt{-2v(z)}d z,
\label{eqn39}
\end{equation}
such that the equation of motion for $\hat{\alpha}$ satisfies
\begin{equation}
\partial' \bar{\partial}' \hat{\alpha}(z',\bar{z}')-\sinh\hat{\alpha}(z',\bar{z}')=0.
\label{eqn322}
\end{equation}

\section{N-spike solution}

We had the fact that the classical strings in $AdS_3$ can be reduced to a generalized sinh-Gordon model coupled to two arbitrary functions $u(\bar{z})$ and $v(z)$ which together represent a free scalar field. These functions are central to the string theory interpretation of the sinh-Gordon equation, they represent the freedom of performing general conformal transformations which are the symmetry of the conformal gauge string.

The general solution to the sinh-Gordon equation (\ref{eqn322}) with N solitons is well known, it can be obtained by the inverse scattering method \cite{Jevicki:2009uz, AKNS} and written in the form
\begin{equation}
\hat{\alpha}(z',\bar{z}')=\ln \Bigl[{4\zeta \over i}{\partial(\varphi_1+\varphi_2) \over \varphi_1-\varphi_2}\Bigr], 
\label{eqn326}
\end{equation}
where $\zeta$ is the spectral parameter and the components of wavefunction $\varphi$ are
\begin{eqnarray}
\varphi_1(\zeta,z',\bar{z}')&=&-\Bigl(\sum_{j,l=1}^N {\lambda_j \over \zeta+\zeta_j}(1-A)_{jl}^{-1}\lambda_l\Bigr)e^{i \zeta \bar{z}'-i z' / 4\zeta}, \label{eqn327} \\
\varphi_2(\zeta,z',\bar{z}')&=&\Bigl(1+\sum_{j,l,k=1}^N {\lambda_j \over \zeta+\zeta_j}{\lambda_j \lambda_l \over \zeta_j+\zeta_l}(1-A)_{lk}^{-1} \lambda_k \Bigr)e^{i \zeta \bar{z}'-i z' / 4\zeta}, \label{eqn328}
\end{eqnarray}
with the definitions
\begin{equation}
A_{ij}=\sum_l a_{il}a_{lj}, \qquad a_{il}={\lambda_i \lambda_l \over \zeta_i+\zeta_l}, \qquad \lambda_k=\sqrt{c_k(0)} e^{i \zeta_k \bar{z}'-i z' / 4 \zeta_k}.
\end{equation}
Here $c_k(0)$ and $\zeta_k$ are two sets of constants related to the initial positions and momenta of N solitons.

The inverse scattering method also gives us a procedure to generate string solutions from the soliton solutions of the sinh-Gordon equation. Essentially the sinh-Gordon configuration serves as a (time-dependent) potential in the Lax scattering equations with scattering wavefunctions. The string coordinates are then directly obtained from these wavefunctions. In this manner the Lax system provides an explicit mapping between the N-soliton sinh-Gordon solutions and the dynamical N-spike AdS string configurations. The fundamental one-to-one correspondence between the solitons/singularities of the sinh-Gordon field and spikes of the string solution was established in detail in our previous work \cite{Jevicki:2007aa}.

We now shortly summarize the construction of the classical string solutions. From the scattering problem associated with the Lax system one finds the scattering wavefunctions
\begin{eqnarray}
\phi_1&=&-{(1+i) \over 2\sqrt{2}}e^{-{i \over 2}(\bar{z}'-z')}\Bigl\{ \sqrt[8]{-v \over u} e^{{1 \over 4}\hat{\alpha}}(\tilde{\varphi}_2-\tilde{\varphi}_1)_-^1 +\sqrt[8]{u \over -v} e^{-{1 \over 4}\hat{\alpha}}(\tilde{\varphi}_2+\tilde{\varphi}_1)_-^1 \Bigr\}, \\
\phi_2&=&+{(1-i) \over 2\sqrt{2}}e^{-{i \over 2}(\bar{z}'-z')}\Bigl\{ \sqrt[8]{-v \over u} e^{{1 \over 4}\hat{\alpha}}(\tilde{\varphi}_2-\tilde{\varphi}_1)_-^1 -\sqrt[8]{u \over -v} e^{-{1 \over 4}\hat{\alpha}}(\tilde{\varphi}_2+\tilde{\varphi}_1)_-^1 \Bigr\},
\end{eqnarray}
where
\begin{equation}
(\tilde{\varphi}_2 \pm \tilde{\varphi}_1)_\pm^1=1 \pm \sum_{j,l}{\lambda_j \over \pm {1 \over 2}+\zeta_j}(1-A)_{jl}^{-1}\lambda_l+\sum_{j,l,k}{\lambda_j \over \pm {1 \over 2}+\zeta_j}{\lambda_j \lambda_l \over \zeta_j+\zeta_l}(1-A)_{lk}^{-1}\lambda_k. \label{eqn363}
\end{equation}
The subscript $\pm$ corresponds to the $\pm$ signs before ${1 \over 2}$.

The second set of linearly independent wavefunction is found to be
\begin{eqnarray}
\psi_1&=&-{1 \over 2\sqrt{2}}\Bigl\{\sqrt[8]{-v \over u} e^{{1 \over 4}\hat{\alpha}}\Bigl[e^{-{1 \over 2}(\bar{z}'+z')}(\tilde{\varphi}_2-\tilde{\varphi}_1)_+^2 +i e^{{1 \over 2}(\bar{z}'+z')}(\tilde{\varphi}_2-\tilde{\varphi}_1)_-^2\Bigr] \cr
& &+i\sqrt[8]{u \over -v} e^{-{1 \over 4}\hat{\alpha}}\Bigl[e^{-{1 \over 2}(\bar{z}'+z')}(\tilde{\varphi}_2+\tilde{\varphi}_1)_+^2 -i e^{{1 \over 2}(\bar{z}'+z')}(\tilde{\varphi}_2+\tilde{\varphi}_1)_-^2 \Bigr] \Bigr\}, \\
\psi_2&=&-{i \over 2\sqrt{2}}\Bigl\{\sqrt[8]{-v \over u} e^{{1 \over 4}\hat{\alpha}}\Bigl[e^{-{1 \over 2}(\bar{z}'+z')}(\tilde{\varphi}_2-\tilde{\varphi}_1)_+^2 +i e^{{1 \over 2}(\bar{z}'+z')}(\tilde{\varphi}_2-\tilde{\varphi}_1)_-^2\Bigr] \cr
& &-i\sqrt[8]{u \over -v} e^{-{1 \over 4}\hat{\alpha}}\Bigl[e^{-{1 \over 2}(\bar{z}'+z')}(\tilde{\varphi}_2+\tilde{\varphi}_1)_+^2 -i e^{{1 \over 2}(\bar{z}'+z')}(\tilde{\varphi}_2+\tilde{\varphi}_1)_-^2 \Bigr] \Bigr\},
\end{eqnarray}
where
\begin{equation}
(\tilde{\varphi}_2 \pm \tilde{\varphi}_1)_\pm^2=1\pm \sum_{j,l}{\lambda_j \over \pm{i \over 2}+\zeta_j}(1-A)_{jl}^{-1}\lambda_l+\sum_{j,l,k}{\lambda_j \over \pm{i \over 2}+\zeta_j}{\lambda_j \lambda_l \over \zeta_j+\zeta_l}(1-A)_{lk}^{-1}\lambda_k.
\end{equation}
Similar to (\ref{eqn363}), the subscript $\pm$ corresponds to the $\pm$ signs before ${i \over 2}$.

The N-spike string solution is given by
\begin{eqnarray}
Z_1&=&{1-i \over 4}e^{{i \over 2}(\bar{z}'-z')}\Bigl\{i(\tilde{\varphi}_2-\tilde{\varphi}_1)_+^1 \Bigl[e^{-{1 \over 2}(\bar{z}'+z')}(\tilde{\varphi}_2+\tilde{\varphi}_1)_+^2 -ie^{{1 \over 2}(\bar{z}'+z')}(\tilde{\varphi}_2+\tilde{\varphi}_1)_-^2 \Bigr] \cr
& &+(\tilde{\varphi}_2+\tilde{\varphi}_1)_+^1 \Bigl[e^{-{1 \over 2}(\bar{z}'+z')}(\tilde{\varphi}_2-\tilde{\varphi}_1)_+^2 +ie^{{1 \over 2}(\bar{z}'+z')}(\tilde{\varphi}_2-\tilde{\varphi}_1)_-^2 \Bigr] \Bigr\},
\label{eqn250} \\
Z_2&=&{1+i \over 4}e^{{i \over 2}(\bar{z}'-z')}\Bigl\{i(\tilde{\varphi}_2-\tilde{\varphi}_1)_+^1 \Bigl[e^{-{1 \over 2}(\bar{z}'+z')}(\tilde{\varphi}_2+\tilde{\varphi}_1)_+^2 +ie^{{1 \over 2}(\bar{z}'+z')}(\tilde{\varphi}_2+\tilde{\varphi}_1)_-^2 \Bigr] \cr
& &+(\tilde{\varphi}_2+\tilde{\varphi}_1)_+^1 \Bigl[e^{-{1 \over 2}(\bar{z}'+z')}(\tilde{\varphi}_2-\tilde{\varphi}_1)_+^2 -ie^{{1 \over 2}(\bar{z}'+z')}(\tilde{\varphi}_2-\tilde{\varphi}_1)_-^2 \Bigr] \Bigr\}.
\label{eqn251}
\end{eqnarray}

Let us summarize the properties of the general solution constructed this way. This general string configuration is characterized by two arbitrary functions $u(\bar{z})$, $v(z)$ and a discrete set of moduli representing the soliton singularities (coordinates). After fixing the conformal frame only the soliton moduli remain giving a specification of the dynamical string moduli. These represent general motions of the spikes and their locations.
 
In general the sinh-Gordon singularities behave as particles and follow interacting particle trajectories. Through our explicit transformations this dynamics translates into the spike dynamics of the $AdS_3$ string. Concretely, given the trajectories of N solitons $x_i(t),i=1,2,\cdots,N$, we can in principle by direct substitution (\ref{eqn250}, \ref{eqn251}) with $\sigma_i(\tau)$ construct the trajectories of N spikes by
\begin{equation}
Z_1^i(\tau)=Z_1(\tau,\sigma_i(\tau)), \qquad Z_2^i(\tau)=Z_2(\tau,\sigma_i(\tau)),
\end{equation}
where $\tau$ acts like the proper time. We therefore have a mapping where on the left hand side the index $i$ labels the string spikes while on the right side it denotes the solitons/singularities. This construction is straightforward in principle, with the map provided by the known wavefunctions of the scattering problem.

\subsection{One-spike dynamical solution}

Now we focus on two simple examples where the energies can be explicitly evaluated. First, for one soliton, we choose the parameters to be
\begin{equation}
u(\bar{z})=2, \quad v(z)=-2, \quad c_1(0)=-2\zeta_1=-i\tilde{v}_1,
\end{equation}
where $\tilde{v}_1 \equiv \sqrt{(1-v_1)/(1+v_1)}$ and $v_1$ is the velocity of soliton on the worldsheet.\footnote{Notice, in general, the worldsheet time $\tau$ is different from the global time $t$.} This will correspond to the one-soliton solution of the sinh-Gordon equation
\begin{equation}
\alpha=\ln\Bigl[2\tanh^2\bigl[{\sigma-v_1\tau \over \sqrt{1-v_1^2}}\bigr]\Bigr].
\end{equation}
In terms of the worldsheet coordinates $\tau$ and $\sigma$, the string solution is given by
\begin{eqnarray}
Z_1&=&{e^{i\tau} \over e^{(\sigma-\tau)/\tilde{v}_1}+e^{-(\sigma+\tau)\tilde{v}_1}}\Bigl\{e^{(\sigma-\tau)/\tilde{v}_1} \cosh\sigma \cr
& &+e^{-(\sigma+\tau)\tilde{v}_1} {(1-i\tilde{v}_1)^2((1+\tilde{v}_1^2)\cosh\sigma+2\tilde{v}_1 \sinh\sigma) \over 1-\tilde{v}_1^4} \Bigr\}, \label{string1} \\
Z_2&=&{-i e^{i\tau} \over e^{(\sigma-\tau)/\tilde{v}_1}+e^{-(\sigma+\tau)\tilde{v}_1}}\Bigl\{e^{(\sigma-\tau)/\tilde{v}_1} \sinh\sigma \cr
& &+e^{-(\sigma+\tau)\tilde{v}_1} {(1-i\tilde{v}_1)^2((1+\tilde{v}_1^2)\sinh\sigma+2\tilde{v}_1 \cosh\sigma) \over 1-\tilde{v}_1^4} \Bigr\}. \label{string2}
\end{eqnarray}
The momenta densities are calculated as
\begin{eqnarray}
{\cal P}_t^\tau={\sqrt{\lambda} \over 2\pi}{\rm Im}(\dot{Z}_1^* Z_1), &\quad& {\cal P}_t^\sigma={\sqrt{\lambda} \over 2\pi}{\rm Im}(Z_1^{'*} Z_1), \\
{\cal P}_\theta^\tau={\sqrt{\lambda} \over 2\pi}{\rm Im}(\dot{Z}_2^* Z_2), &\quad& {\cal P}_\theta^\sigma={\sqrt{\lambda} \over 2\pi}{\rm Im}(Z_2^{'*} Z_2),
\end{eqnarray}
where $\lambda$ is the coupling constant. One can explicitly check ${\cal P}_t^\sigma$ and ${\cal P}_\theta^\sigma$ vanish asymptotically at $\sigma=\pm \infty$ as long as the string solution is regular. That is, in the singular case when $v_1=0$, we found nonvanishing momentum flow with $\tau$ dependence at the boundary of the string \cite{Jevicki:2007aa}. 

By the current conservation $\partial_\tau {\cal P}_{t,\theta}^\tau-\partial_\sigma {\cal P}_{t,\theta}^\sigma=0$, we can calculate the energy and angular momentum at any convenient $\tau$. For instance, the energy and momentum densities at $\tau=0$ are simplified to be
\begin{eqnarray}
{\cal P}_t^\tau &=& {\sqrt{\lambda} \over 2\pi(e^{\sigma/\tilde{v}_1}+e^{-\sigma\tilde{v}_1})^2}\Bigl\{e^{2\sigma/\tilde{v}_1}\cosh^2\sigma+e^{-2\sigma\tilde{v}_1}\Bigl({\cosh\sigma+\epsilon_1^{-1}\sinh\sigma \over v_1}\Bigr)^2\Bigr\}, \\
{\cal P}_\theta^\tau &=& {\sqrt{\lambda} \over 2\pi(e^{\sigma/\tilde{v}_1}+e^{-\sigma\tilde{v}_1})^2}\Bigl\{e^{2\sigma/\tilde{v}_1}\sinh^2\sigma+e^{-2\sigma\tilde{v}_1}\Bigl({\sinh\sigma+\epsilon_1^{-1}\cosh\sigma \over v_1}\Bigr)^2\Bigr\},
\end{eqnarray}
where $\epsilon_1 \equiv (1-v_1^2)^{-1/2}$ is the energy of the soliton.

Introduce a cutoff $\Lambda$ for the $\sigma$ integration, up to the subleading term, the energy and angular momentum are
\begin{eqnarray}
E&=&\int_{-\Lambda}^\Lambda {\cal P}_t^\tau d\sigma \sim {\sqrt{\lambda} \over 2\pi} \Bigl[ {1 \over 4(1+\epsilon_1^{-1})} e^{2\Lambda}+\Lambda \Bigr], \\
S&=&\int_{-\Lambda}^\Lambda {\cal P}_\theta^\tau d\sigma \sim {\sqrt{\lambda} \over 2\pi} \Bigl[ {1 \over 4(1+\epsilon_1^{-1})} e^{2\Lambda}-\Lambda \Bigr].
\end{eqnarray}
Therefore, the difference between $E$ and $S$ can be calculated as
\begin{equation}
E-S={\sqrt{\lambda} \over 2\pi}\int_{-\Lambda}^\Lambda {\cosh[2\epsilon_1\sigma] \over 1+\cosh[2\epsilon_1\sigma]}d\sigma \sim {\sqrt{\lambda} \over 2\pi} \Bigl[\ln{8\pi S \over \sqrt{\lambda}}+\ln(1+\epsilon_1^{-1})-\epsilon_1^{-1}\Bigr].
\label{dispersion1}
\end{equation}
Measuring the energy from the infinite GKP string, the dispersion relation can be written as \cite{Dorey:2010iy}
\begin{equation}
E-S=E_0 + {\sqrt{\lambda} \over 2\pi}\Bigl[ {1 \over 2}\ln{1+\epsilon_1^{-1} \over 1-\epsilon_1^{-1}}-\epsilon_1^{-1} \Bigr]
\label{dispersion2}
\end{equation}
with the excitation energy
\begin{equation}
E_{\rm spike}^1(\epsilon_1) \equiv E-S-E_0={\sqrt{\lambda} \over 2\pi}\Bigl[ {1 \over 2}\ln{1+\epsilon_1^{-1} \over 1-\epsilon_1^{-1}}-\epsilon_1^{-1} \Bigr].
\label{spikeenergy}
\end{equation}
The inverse power in the soliton energy shows a similarity with the case of giant magnons on $R \times S^2$ \cite{Hofman:2006xt} where the excitation energy of the string (and the giant magnon) is equal to the inverse of the sine-Gordon soliton energy.\footnote{There one uses a time-like gauge $t=\tau$.} Here in the AdS case we find extra logarithmic terms pointing to a more complex dynamical system governing the spike dynamics in AdS as compared with $R \times S^2$.

\subsection{Two-spike dynamics}

Our next explicit example is for two solitons, where the parameters are chosen to be
\begin{equation}
c_1(0)=2\zeta_1 {\zeta_1+\zeta_2 \over \zeta_1-\zeta_2}, \quad c_2(0)=2\zeta_2 {\zeta_1+\zeta_2 \over \zeta_1-\zeta_2}, \quad \zeta_1=-{i \over 2}\tilde{v}_1, \quad \zeta_2={i \over 2}\tilde{v}_2^{-1},
\end{equation}
where $\tilde{v}_{1,2} \equiv \sqrt{(1-v_{1,2})/(1+v_{1,2})}$ and $v_{1,2}$ are the magnitudes of the velocities of the two solitons moving towards each other. The special case $v_1=v_2=v$ corresponds to the two-soliton solution used in \cite{Jevicki:2007aa}
\begin{equation}
\alpha=\ln\Bigl[2\Bigl({v\cosh X-\cosh T \over v\cosh X+\cosh T}\Bigr)^2\Bigr],
\end{equation}
where $X \equiv 2\gamma\sigma,T \equiv 2v\gamma\tau$.

The exact expressions for the momenta densities are lengthy to write them down, but the leading terms at $\tau=0$ are
\begin{eqnarray}
{\cal P}_t^\tau &\sim& {\sqrt{\lambda} \over 2\pi}\Bigl[ {1 \over 4}\Bigl({1-\tilde{v}_1 \over 1+\tilde{v}_1}\Bigr)^2 e^{2\sigma}+{1 \over 4}\Bigl({1-\tilde{v}_2 \over 1+\tilde{v}_2}\Bigr)^2 e^{-2\sigma}+{1 \over 2}\Bigr], \\
{\cal P}_\theta^\tau &\sim& {\sqrt{\lambda} \over 2\pi}\Bigl[ {1 \over 4}\Bigl({1-\tilde{v}_1 \over 1+\tilde{v}_1}\Bigr)^2 e^{2\sigma}+{1 \over 4}\Bigl({1-\tilde{v}_2 \over 1+\tilde{v}_2}\Bigr)^2 e^{-2\sigma}-{1 \over 2}\Bigr].
\end{eqnarray}
In the special case where $v_1=v_2=v$, the dispersion relation is
\begin{equation}
E-S = {\sqrt{\lambda} \over 2\pi}\Bigl[ \ln{8\pi S \over \sqrt{\lambda}}+\ln{1+\epsilon^{-1} \over 1-\epsilon^{-1}}-2\epsilon^{-1}+\cdots \Bigr].
\end{equation}
For different velocities, using the symmetric $\rho$ regularization, we obtain the result
\begin{equation}
E-S={\sqrt{\lambda} \over 2\pi}\Bigl[ \ln{8\pi S \over \sqrt{\lambda}}+{1 \over 2}\ln{1+\epsilon_1^{-1} \over 1-\epsilon_1^{-1}}+{1 \over 2}\ln{1+\epsilon_2^{-1} \over 1-\epsilon_2^{-1}}-\epsilon_1^{-1}-\epsilon_2^{-1} \Bigr].
\end{equation}
Following the definition of spike energy (\ref{spikeenergy}), the excitation energy of the two-spike solution is given by the sum of two individual spike energies
\begin{equation}
E_{\rm spike}^2(\epsilon_1,\epsilon_2)=E_{\rm spike}^1(\epsilon_1)+E_{\rm spike}^1(\epsilon_2).
\end{equation}
In the center of mass frame $v_1=v_2=v$, the energy is $E_{\rm spike}^2(\epsilon)=2E_{\rm spike}^1(\epsilon)$. For this special case, the result is also obtained in \cite{Dorey:2010iy}.

\section{Solitons and Cusps: Euclidean worldsheet}

We have described in Introduction the relevance of Euclidean worldsheet solution. The simplest example of such a solution is the cusp near the boundary of AdS. We describe how this solution can be simply found from the one-soliton solution of sinh-Gordon using an Euclidean version of the inverse scattering method.

Start with $AdS_3$ metric in Poincar\'{e} coordinates
\begin{equation}
ds^2={-d y_0^2+d y_1^2+d r^2 \over r^2},
\end{equation}
the single cusp solution sitting at the boundary $r=0$ can be found using the ansatz
\begin{equation}
y_0=e^s \cosh \theta, \quad y_1=e^s \sinh \theta; \quad r=e^s w(s).
\label{cusp}
\end{equation}
One can check explicitly that $w(s)=\sqrt{2}$ solves the equation of motion \cite{Alday:2007hr}. This single cusp solution can be regularized to some finite boundary at $r=r_c$ \cite{Berkovits:2008ic} with the same ansatz (\ref{cusp}) while $w(s)$ is now implicitly given by the equation
\begin{equation}
{e^s \over r_c}=\Bigl( {w+\sqrt{2} \over w-\sqrt{2}} \Bigr)^{1/\sqrt{2}} {1 \over 1+w}.
\end{equation}
The cusp is approached at $s \to -\infty$. In many circumstances, the conformal gauge solutions are desirable \cite{Alday:2009zm, Alday}. Choosing the ansatz
\begin{equation}
y_0=f(\tau)\cosh\sigma, \quad y_1=f(\tau)\sinh\sigma; \quad r=g(\tau),
\label{solreg}
\end{equation}
one finds that the functions $f(\tau)$ and $g(\tau)$ are solved to be
\begin{eqnarray}
f(\tau)&=&\sqrt{2} e^\tau r_c {\tanh [\tau/\sqrt{2}] \over 2+\sqrt{2}\tanh [\tau/\sqrt{2}]}, \\
g(\tau)&=&2 e^\tau r_c {1 \over 2+\sqrt{2}\tanh [\tau/\sqrt{2}]},
\end{eqnarray}
where the cusp is approached at $\tau=0$.

To develop the general method for constructing AdS string solutions in conformal gauge, we employ the technique developed in \cite{Jevicki:2007aa, Alday:2009yn} based on Pohlmeyer reduction and inverse scattering. Now for Euclidean worldsheets, all one needs to change is $\tau \to -i\tau$. In the case of $AdS_3$, the reduction map is given in terms of the embedding coordinates by
\begin{eqnarray}
e^{2\alpha(z,\bar{z})}&=&{1 \over 2} \partial Y \cdot \bar{\partial} Y, \\
B_a &=& {e^{-2\alpha} \over 2}\epsilon_{abcd} Y^b \partial Y^c \bar{\partial} Y^d, \\
p&=&{1 \over 2} B \cdot \partial^2 Y, \quad \bar{p}=-{1 \over 2} B \cdot \bar{\partial}^2 Y.
\end{eqnarray}
The equation of motion for the scalar field $\alpha(z,\bar{z})$ turns out to be
\begin{equation}
\partial \bar{\partial} \alpha(z,\bar{z}) - e^{2\alpha} + p(z)\bar{p}(\bar{z}) e^{-2\alpha}=0,
\end{equation}
with a holomorphic function $p(z)$.

In order to construct the string solution, one needs to solve two SL(2) scattering problems. Each of these problems has two linearly independent solutions $\psi_{\alpha,a}^L$, $a=1,2$ and $\psi_{\dot{\alpha},\dot{a}}^R$, $\dot{a}=1,2$ which are normalized as
\begin{eqnarray}
\psi_a^L \wedge \psi_b^L &\equiv& \epsilon^{\beta\alpha}\psi_{\alpha,a}^L \psi_{\beta,b}^L=\epsilon_{ab}, \label{norm1} \\
\psi_{\dot{a}}^R \wedge \psi_{\dot{b}}^R &\equiv& \epsilon^{\dot{\beta}\dot{\alpha}}\psi_{\dot{\alpha},\dot{a}}^R \psi_{\dot{\beta},\dot{b}}^R=\epsilon_{\dot{a}\dot{b}}. \label{norm2}
\end{eqnarray}
The string solution can be written in terms of a combination of the wavefunctions
\begin{equation}
Y_{a\dot{a}}=\begin{pmatrix} Y_{-1}+Y_2 & Y_1-Y_0 \cr Y_1+Y_0 & Y_{-1}-Y_2 \end{pmatrix}_{a\dot{a}}=\psi_{\alpha,a}^L M_1^{\alpha \dot{\beta}} \psi_{\dot{\beta},\dot{a}}^R, \qquad M_1^{\alpha\dot{\beta}}=\begin{pmatrix} 1 & 0 \cr 0 & 1 \end{pmatrix}.
\end{equation}

In a series of papers \cite{Jevicki:2007aa, Jevicki:2008mm, Jevicki:2009uz}, it is shown that solitons in the field theory correspond to cusps of the string. For the construction of the single cusp string solution (\ref{solreg}), one uses the one-soliton solution of the sinh-Gordon equation
\begin{equation}
\alpha(z,\bar{z})={1 \over 2}\ln\Bigl[{1 \over 2}\tanh^2 {\tau \over \sqrt{2}}\Bigr],
\end{equation}
together with the holomorphic function $p(z)=i/2$.

The Dirac wavefuntions are solved to be
\begin{eqnarray}
\Omega^L&=&\begin{pmatrix} \psi_{1,1}^L & \psi_{1,2}^L \cr \psi_{2,1}^L & \psi_{2,2}^L \end{pmatrix}=\begin{pmatrix} \psi_1^L & \bar{\psi_2^L} \cr \psi_2^L & \bar{\psi_1^L} \end{pmatrix}, \\
\Omega^R&=&\begin{pmatrix} \psi_{1,1}^R & \psi_{1,2}^R \cr \psi_{2,1}^R & \psi_{2,2}^R \end{pmatrix}=\begin{pmatrix} \psi_1^R & \bar{\psi_2^R} \cr \psi_2^R & \bar{\psi_1^R} \end{pmatrix},
\end{eqnarray}
where
\begin{eqnarray}
\psi_1^L={e^{i {5\pi \over 8}}\bigl( c_1 e^{{\sigma-\tau \over 2}} \bigl( (\sqrt{2}-1)e^{-{\tau \over \sqrt{2}}}-i e^{\tau \over \sqrt{2}} \bigr)-c_2 e^{-{\sigma-\tau \over 2}} \bigl( (\sqrt{2}+1) e^{-{\tau \over \sqrt{2}}}+i e^{\tau \over \sqrt{2}} \bigr) \bigr) \over 2 \sqrt{\sinh[\sqrt{2}\tau]}}, \\
\psi_2^L={e^{-i {\pi \over 8}}\bigl( c_1 e^{{\sigma-\tau \over 2}} \bigl( (\sqrt{2}-1)e^{-{\tau \over \sqrt{2}}}+i e^{\tau \over \sqrt{2}} \bigr)+c_2 e^{-{\sigma-\tau \over 2}} \bigl( (\sqrt{2}+1) e^{-{\tau \over \sqrt{2}}}-i e^{\tau \over \sqrt{2}} \bigr) \bigr) \over 2 \sqrt{\sinh[\sqrt{2}\tau]}}, \\
\psi_1^R={e^{i {5\pi \over 8}}\bigl( c_3 e^{{\sigma+\tau \over 2}} \bigl( (\sqrt{2}+1)e^{-{\tau \over \sqrt{2}}}+i e^{\tau \over \sqrt{2}} \bigr)-c_4 e^{-{\sigma+\tau \over 2}} \bigl( (\sqrt{2}-1) e^{-{\tau \over \sqrt{2}}}-i e^{\tau \over \sqrt{2}} \bigr) \bigr) \over 2 \sqrt{\sinh[\sqrt{2}\tau]}}, \\
\psi_2^R={e^{-i {\pi \over 8}}\bigl( c_3 e^{{\sigma+\tau \over 2}} \bigl( (\sqrt{2}+1)e^{-{\tau \over \sqrt{2}}}-i e^{\tau \over \sqrt{2}} \bigr)+c_4 e^{-{\sigma+\tau \over 2}} \bigl( (\sqrt{2}-1) e^{-{\tau \over \sqrt{2}}}+i e^{\tau \over \sqrt{2}} \bigr) \bigr) \over 2 \sqrt{\sinh[\sqrt{2}\tau]}}.
\end{eqnarray} 
The normalization conditions (\ref{norm1}-\ref{norm2}) lead to
\begin{eqnarray}
\det \Omega^L&=&c_1^* c_2+c_1 c_2^*=1, \\
\det \Omega^R&=&c_3^* c_4+c_3 c_4^*=1.
\end{eqnarray}
One can choose the constants to be
\begin{equation}
c_1=-c_4=\Bigl({\sqrt{2}+1 \over 2 r_c}\Bigr)^{1/2}, \qquad c_2=-c_3=\Bigl({(\sqrt{2}-1)r_c \over 2}\Bigr)^{1/2},
\end{equation}
such that the string solution
\begin{equation}
(\Omega^L)^T \Omega^R=\begin{pmatrix} Y_{-1}+i Y_0 & Y_1+i Y_2 \cr Y_1-i Y_2 & Y_{-1}-i Y_0 \end{pmatrix},
\end{equation}
reduces to
\begin{equation}
Y_{a\dot{a}}=\begin{pmatrix} {1 \over 2e^\tau r_c}(2+\sqrt{2}\tanh{\tau \over \sqrt{2}}) & {1 \over \sqrt{2}}e^{\sigma}\tanh{\tau \over \sqrt{2}} \cr -{1 \over \sqrt{2}}e^{-\sigma}\tanh{\tau \over \sqrt{2}} & -{e^\tau r_c \over 2}(-2+\sqrt{2}\tanh{\tau \over \sqrt{2}}) \end{pmatrix},
\end{equation}
which agrees with (\ref{solreg}) in terms of Poincar\'{e} coordinates.

\section{Series Method of the AdS$_4$ system}

For the problem of constructing minimal area surfaces in AdS one requires more general solutions with Euclidean worldsheet. At the level of Pohlmeyer reduced Toda field theories these solutions correspond to instanton-type configurations. The inverse scattering techniques were very successful for a general construction of multi-soliton solutions but do not apply for the multi-instanton case. We have for that problem formulated an approximate technique in \cite{Jevicki:2009bv}, which is based on an asymptotic expansion at large distances with matching at short distances. The construction is nontrivial due to the fact that the required instanton type solutions come with specified singularities at the origin.

In this case of $AdS_4$, we have two sets of reduced fields $u_i$ and $v_i$ which are parameterized as
\begin{eqnarray}
u_4=-\bar{p}(\bar{z}) \cos \gamma(z,\bar{z}), &\quad& v_4=+p(z) \cos \bar{\gamma}(z,\bar{z}), \\
u_5=-\bar{p}(\bar{z}) \sin \gamma(z,\bar{z}), &\quad& v_5=-p(z) \sin \bar{\gamma}(z,\bar{z}).
\end{eqnarray}
Defining a new field $\beta(z,\bar{z}) \equiv \gamma(z,\bar{z})+\bar{\gamma}(z,\bar{z})$, the zero curvature condition implies the equations of motion for the scalar fields
\begin{eqnarray}
&&\partial \bar{\partial} \alpha-e^\alpha+p(z)\bar{p}(\bar{z})e^{-\alpha} \cos \beta=0, \label{eomlate1} \\
&&\partial \bar{\partial} \beta-p(z)\bar{p}(\bar{z})e^{-\alpha} \sin \beta=0. \label{eomlate2}
\end{eqnarray}
After a shift of scalar field $\alpha(z,\bar{z})=\hat{\alpha}(z,\bar{z})+{1 \over 2}\ln[p(z)\bar{p}(\bar{z})]$, and a change of variables $dw=\sqrt{p(z)}dz$, the equations of motion (\ref{eomlate1}-\ref{eomlate2}) become
\begin{eqnarray}
&&\partial_w \bar{\partial}_{\bar{w}} \hat{\alpha}-e^{\hat{\alpha}}+e^{-\hat{\alpha}}\cos\beta=0, \label{toda1} \\
&&\partial_w \bar{\partial}_{\bar{w}} \beta-e^{-\hat{\alpha}}\sin\beta=0. \label{toda2}
\end{eqnarray}
This is the $B_2$ Toda system of equations.

The essence of the method that we will be following is to perform an expansion at large distance and then impose the short distance boundary condition (in this case the singularity condition). Even though this represents an extrapolation of the series solution from large all the way to small distance, the technique was known to give excellent results in the case of nonlinear soliton solutions.

We impose the boundary condition for the Toda system at infinity
\begin{equation}
\hat{\alpha},\beta \to 0 \qquad {\rm at}~~\rho=\infty,
\label{bc}
\end{equation}
and generate the large distance expansion with the form
\begin{equation}
\hat{\alpha}=\sum_{n=1}^\infty \alpha_n, \qquad \beta=\sum_{n=1}^\infty \beta_n.
\end{equation}
The first order solutions satisfying the boundary condition (\ref{bc}) are
\begin{eqnarray}
&&\alpha_1(\tilde{\rho})=a K_0(\tilde{\rho}), \\
&&\beta_1(\rho)=b K_0(\rho),
\end{eqnarray}
where $\tilde{\rho} \equiv \sqrt{2}\rho$ and $a,b$ are integration constants. The second order solutions are written in terms of double integrals
\begin{eqnarray}
&&\alpha_2(\tilde{\rho})=+K_0(\tilde{\rho}) \int_{\tilde{\rho}}^\infty {d\tilde{\rho}' \over \tilde{\rho}' K_0^2(\tilde{\rho}')} \int_{\tilde{\rho}'}^\infty {1 \over 4}b^2 K_0^2(\tilde{\rho}'' / \sqrt{2}) K_0(\tilde{\rho}'') \tilde{\rho}'' d\tilde{\rho}'' , \\
&&\beta_2(\rho)=-K_0(\rho)\int_\rho^\infty {d\rho' \over \rho' K_0^2(\rho')} \int_{\rho'}^\infty a b K_0(\sqrt{2} \rho'')K_0^2(\rho'') \rho'' d\rho'' .
\end{eqnarray}
The third order solutions are
\begin{eqnarray}
&&\alpha_3(\tilde{\rho})=+K_0(\tilde{\rho}) \int_{\tilde{\rho}}^\infty {d\tilde{\rho}' \over \tilde{\rho}' K_0^2(\tilde{\rho}')} \int_{\tilde{\rho}'}^\infty S_3^\alpha(\tilde{\rho}'') K_0(\tilde{\rho}'') \tilde{\rho}'' d\tilde{\rho}'' , \\
&&\beta_3(\rho)=-K_0(\rho)\int_\rho^\infty {d\rho' \over \rho' K_0^2(\rho')} \int_{\rho'}^\infty S_3^\beta(\rho'') K_0(\rho'') \rho'' d\rho'' ,
\end{eqnarray}
where the source terms are
\begin{eqnarray}
&&S_3^\alpha(\tilde{\rho}'')={1 \over 6}\alpha_1^3-{1 \over 4}\alpha_1\beta_1^2+{1 \over 2}\beta_1 \beta_2 \Bigr|_{\tilde{\rho}''}, \\
&&S_3^\beta(\rho'')={1 \over 6}\beta_1^3-{1 \over 2}\alpha_1^2 \beta_1+\alpha_1\beta_2+\alpha_2\beta_1 \Bigr|_{\rho''}.
\end{eqnarray}

As pointed out in \cite{Alday:2009dv}, it is relevant to continue from the Lorentzian AdS space with (1,3) signature to (2,2) signature. The boundary conditions are different in these two cases. For AdS$_4$, this analytic continuation can be achieved by taking $\beta \to i\beta$ of (\ref{toda1}-\ref{toda2}) and one has
\begin{eqnarray}
&&\partial_w \bar{\partial}_{\bar{w}}\hat{\alpha}-e^{\hat{\alpha}}+e^{-\hat{\alpha}}\cosh\beta=0, \label{eom3} \\
&&\partial_w \bar{\partial}_{\bar{w}}\beta-e^{-\hat{\alpha}}\sinh\beta=0. \label{eom4}
\end{eqnarray}
If we assume an expansion of the embedding coordinates
\begin{equation}
Y_{-2}=1+\sum_{l=1} \Phi_l^{(-2)} (\varphi) \rho^l, \quad Y_i=\sum_{l=1} \Phi_l^{(i)} (\varphi) \rho^l, \quad i=-1,0,1,2,
\end{equation}
where $\rho,\varphi$ are the polar worldsheet coordinates $z=\rho e^{i\varphi},\bar{z}=\rho e^{-i\varphi}$. The string sigma model (\ref{stringsigma}) and Virasoro constraints (\ref{Virasoro}) give
\begin{equation}
[\cosh\beta ~~{\rm or}~ \cos\beta]={C_0+C_z z+C_{\bar{z}} \bar{z}+\cdots \over \rho^{2n-4}},
\end{equation}
where we have used the polynomial form of the holomorphic function $p(z)=z^{n-2}$. In the case of (2,2) signature, as $\cosh\beta \ge 1$, we expect $C_0 \neq 0$ as compared with the (1,3) signature where $\cos\beta \le 1$, so one expects $C_0=0$.

To summarize, in the case of $(1,3)$ signature, only the $\hat{\alpha}$ field has singularity
\begin{equation}
\hat{\alpha}=-2\zeta\ln\rho+c_\alpha+\cdots, \qquad \beta=c_\beta+\cdots,
\label{bcnew1}
\end{equation}
while in the case of $(2,2)$ signature, both scalar fields $\hat{\alpha}$ and $\beta$ can have logarithmic singularities near $\rho=0$ as
\begin{equation}
\hat{\alpha}=-2\zeta\ln\rho+c_\alpha+\cdots, \qquad \beta=-4\zeta\ln\rho+c_\beta+\cdots.
\end{equation}
As for the large $\rho$ solution, all one needs to do is to replace $b \to i b$ in the expressions above. Now the matching conditions for the singularities give
\begin{eqnarray}
a-{\pi \over 8}b^2+{\pi^2 \over 96}a^3+{\pi^2 \over 32}a b^2&=&2{n-2 \over n}, \\
b-{\pi \over 8}a b+{7\pi^2 \over 384}b^3+{3\pi^2 \over 128}a^2 b&=&4{n-2 \over n},
\end{eqnarray}
and the area is given by
\begin{eqnarray}
A_{Toda}^{++}&=&{\pi n \over 4}\int_0^\infty \tilde{\rho} d\tilde{\rho} (\alpha_1+\alpha_2+{1 \over 2}\alpha_1^2+\alpha_3+\alpha_1\alpha_2+{1 \over 6}\alpha_1^3+\cdots), \\
&=&{\pi n \over 4}(a+{1 \over 4}a^2-0.143 b^2+0.103 a^3-0.0843 a b^2+\cdots).
\end{eqnarray}
In the case of pentagon where $2n=5$, we find $A_{\rm pentagon}^{++} \approx 0.97849$. Comparing with the shooting method result \cite{Alday:2009dv}, the difference is about 17\%. Better accuracy can be obtained by adding more terms.

\section*{Acknowledgments}

We would like to thank J. Avan, C. Kalousios and C. Vergu for comments and discussions. This work is supported by the Department of Energy under contract DE-FG02-91ER40688. The work of KJ is also supported by the Galkin fellowship at Brown University.

\end{document}